# Resonant Precession Modulation Based Magnetic Field Receivers

Kevin Quy Tanh Luong, Wei Gu, Foad Fereidoony, Lap Yeung, Zhi Yao & Yuanxun Ethan Wang

## ABSTRACT

Antenna technology provides the crucial interface between electronic devices and electromagnetic waves required for wireless communications. This technology has operated under largely the same general principles for over a century, resulting in stagnation of advancements in modern times and fundamentally limiting the extent to which potential applications of wireless communications may be realized. Herein we propose a novel concept for antenna operation employing the nonlinear precession dynamics of subatomic spins. An electromagnetic wave incident on a precessing spin will serve to modulate the spin frequency of precession. Antennas operating based on a detection of these spin dynamics are then able to fully characterize the incident wave while achieving substantial amplifications, higher sensitivities, and smaller form factors as compared to conventional antennas. A preliminary experimental implementation of the proposed concept establishes its feasibility and is already able to demonstrate significant amplification and size benefits.

## INTRODUCTION

The spin dynamics of subatomic particles serve as the foundation upon which some of the most revolutionary tools and methods of scientific study are based. A partial list of these includes magnetic resonance imaging (MRI), one of the most ubiquitous and safest means of medical imaging available today [1,2], magnetic resonance spectroscopy (MRS), arguably the most popular means of studying molecular structure in chemistry [3,4], and electron spin resonance (ESR) spectroscopy, long the standard technique for the detection, identification, and characterization of free radicals in biology [5]. While spanning a diverse range of applications, all of these groundbreaking tools have in common a dependence on a single feature of spin dynamics—its capability to characterize an environment with remarkable detail and precision. Such a capability has not been overlooked as a means of characterizing environmental magnetic fields in the domain of magnetometry [6,7] where spin dynamics-based approaches make up the sub-category of optical magnetometry and have enabled systems unrivaled in many aspects of performance [8,9]. For example, spin exchange relaxation free (SERF) optical magnetometers have sensitivities surpassing those of any other system [10,11], and nitrogen vacancy (NV) center optical magnetometers are currently of great interest for their potential to realize measurements with unprecedented resolutions [12,13].

The success of spin dynamics-based magnetic field characterization, though undisputable with optical magnetometry, is nevertheless narrow in scope. Optical magnetometers, operating at the limits of achievable performance, are typically suitable for use only as research tools or in specialized niche areas [7,8]; however, the general practice of magnetic field characterization is much more extensive. Particularly prevalent is its application in the domain of wireless communications in order to analyze electromagnetic waves and ultimately extract the information they carry. The effectiveness by which this analysis may be performed depends largely on how well a magnetic field may be transduced to an electrical signal, a feat accomplished with what is known in this domain as a receiving antenna. For over a century [14], receiving antennas have for the most part been operating exclusively under a principle of directly induced voltages on a conducting structure [15,16] and are now a mature technology seeing only nominal advancements. Regardless, they represent a vital part of any wireless system and, with the strong continued growth of the already huge wireless communications market [17] and constant emergence of new wireless technologies [18], improved receiving antenna performance is always sought after. To meet this demand in modern times, it is necessary to deviate from the conventional principles of receiving antenna operation.

The transduction performance of a receiving antenna is strongly correlated to its field sensitivity of which it is known that optical magnetometers are second to none. Unfortunately, whereas it tends to be a requirement that receiving antennas be relatively inexpensively, portable, and readily manufactured [17], optical magnetometers are in contrast large and expensive in terms of both manufacturing and power consumption [7]. Nevertheless, the underlying capabilities of spin dynamics upon which operation of optical magnetometers is predicated have well proven advantages and, with a proper approach, these can be harnessed to introduce a new paradigm of operation for receiving antennas. In this work, we propose for the first time an extension of the concept of spin dynamics-based magnetic field characterization to receiving antennas for wireless communications using a method of resonant precession modulation (RPM). This method relies upon nonlinear spin precession dynamics and is shown to allow for receiving antennas with significant amplifications, higher sensitivities, as well as much more compact sizes as compared to receiving antennas operating under conventional principles. The following sections will cover the basic principles of the method, develop analytical expressions describing it, discuss its advantages and the approach taken to practically implement it, and finally present results from simulations and a preliminary experiment.

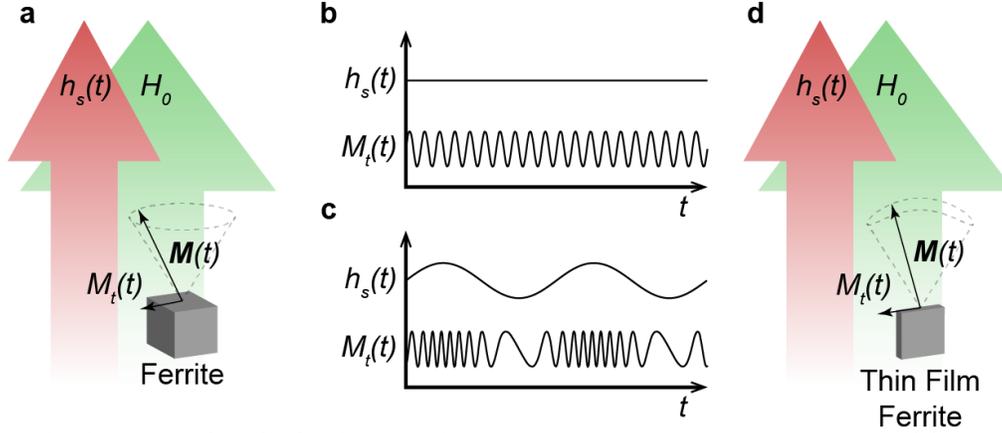

**Fig 1. Magnetization dynamics of a ferrite material. a** Illustration of ferrite magnetization $M(t)$ precession in the presence of a magnetic bias field with both a static $H_0$ and time varying $h_s(t)$ component. **b** Time domain plot qualitatively depicting $h_s(t)$ as well as a component of the ferrite magnetization $M_t(t)$ transverse to the bias field for the case where $h_s(t)$ is constant. **c** Similar plot depicting both $h_s(t)$ and $M_t(t)$ for the case where $h_s(t)$ is varying sinusoidally. **d** Illustration of ferrite magnetization precession for a thin film ferrite in the presence of a magnetic bias field. Demagnetization effects cause the precession to resemble a pendulum oscillation.

## RESULTS

**Basic principles of operation.** Although inspired by optical magnetometry, an extension of spin dynamics-based magnetic field characterization to receiving antennas must be accompanied with some major changes to rectify those aspects which contribute to the unsuitability of optical magnetometers for use in wireless communications. Foremost among these changes is with the medium from which the spins are sourced. Optical magnetometers use spins sourced from atomic vapors [8,9], the glass containers of which are the bottleneck preventing reductions in cost and size of the device [7]. In contrast, we focus instead on spins sourced from insulating ferrimagnetic materials, or ferrites [19,20]. The insulating nature, low magnetic loss at high frequencies, and decent saturation magnetization values [21] of ferrites will all prove imperative to achieving a high performance receiving antenna. It is in fact these same attributes that have motivated past antennas incorporating ferrites [15,16,22,23,24], though none have ever employed the use of nonlinear spin precession dynamics as is done in this work. Ferrites additionally have an extensive history of incorporation in general high frequency electronics [19,20] resulting in now well-established processing methods [21] that allow the material to be produced relatively inexpensively.

With ferrites as the medium of choice, analysis of spin dynamics may be accomplished through application of micromagnetic theory [25]. This theory is able to practically describe the average behavior of the large ensemble of spins of a macroscopic ferrite material [26,27] and is formulated classically in terms of the spatially and temporally dependent material magnetization $M(r,t)$ based on a continuum approximation [25]. References to magnetization as opposed to spin will consequently be made for all following analyses for the purposes of consistency, where it is recognized that either may be immediately determined based on knowledge of the other [25]. From micromagnetic theory, the magnetization dynamics of a ferrite material are governed by the Landau-Lifshitz-Gilbert (LLG) equation [28]

$$\frac{d\boldsymbol{M}(\boldsymbol{r},t)}{dt} = -\gamma\mu_0(\boldsymbol{M}(\boldsymbol{r},t) \times \boldsymbol{H}_e(\boldsymbol{r},t)) + \frac{\alpha}{M_s}\left(\boldsymbol{M}(\boldsymbol{r},t) \times \frac{d\boldsymbol{M}(\boldsymbol{r},t)}{dt}\right) \quad (1)$$

where $\gamma$ is the gyromagnetic ratio, $\mu_0$ is the permeability of free space, $\boldsymbol{H}_e(\boldsymbol{r},t)$ is an effective magnetic field representing actual magnetic fields as well as various other physics, $\alpha$ is the Gilbert damping coefficient representing magnetic loss, and $M_s = |\boldsymbol{M}(\boldsymbol{r},t)|$ is the saturation magnetization. For the purposes of simplicity, the assumption will be made initially that $\boldsymbol{H}_e(\boldsymbol{r},t)$ is spatially uniform with the dominant contributor to its value being some constant magnetic field $\boldsymbol{H}_0$. It will further be assumed that the magnitude $|\boldsymbol{H}_e(\boldsymbol{r},t)| = H_e(\boldsymbol{r},t)$ is sufficiently large such that $\boldsymbol{M}(\boldsymbol{r},t)$ is also spatially uniform, and that magnetic loss is neglected.

From the LLG equation (1), the means by which nonlinear spin precession dynamics may be used to characterize magnetic fields of electromagnetic waves is elucidated. The equation indicates the equilibrium state of $M(t)$ to be aligned with $H_e$, where it is noted that spatial and temporal dependencies of these and all following variables will be omitted as appropriate. Any perturbation from the equilibrium state will result in a continuous precession of $M(t)$ about $H_e$, visualized in Fig. 1a and b, at the resonance frequency $f_0$ given by

$$f_0 = \frac{1}{2\pi}\gamma\mu_0 H_e. \tag{2}$$

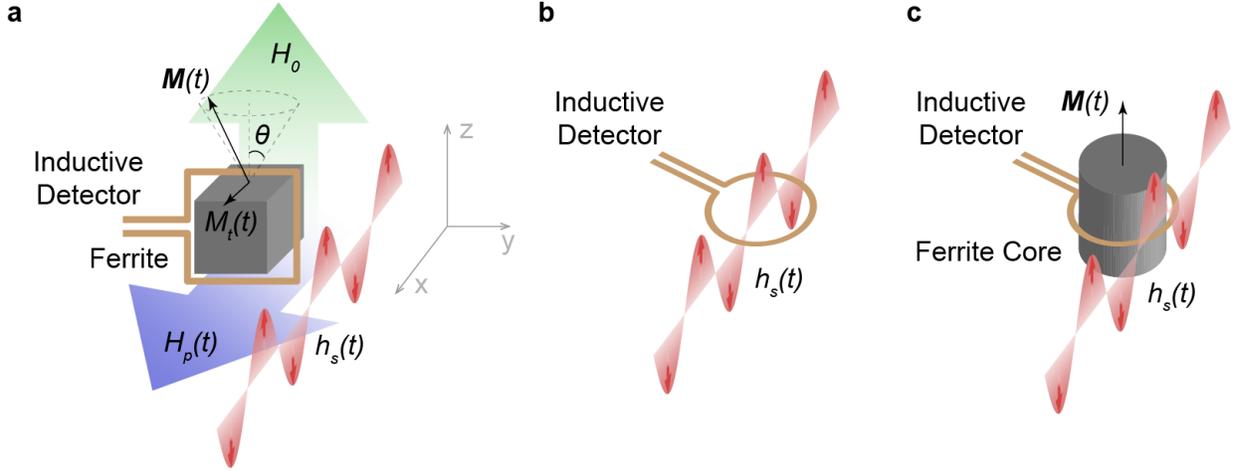

**Fig. 2. Receiving antenna operation. a** Illustration of receiving antenna operation under the principles of RPM where orientations of the bias magnetic field $H_0$, the perturbing magnetic field $H_p(t)$, the incident wave magnetic field $h_s(t)$, and the inductive detector are visualized with respect to the ferrite. **b** Illustration of receiving antenna operation under conventional principles. **c** Illustration of ferrite rod receiving antenna operation where positioning of the ferrite core within the inductive detector is visualized.

If magnetic loss is accounted for, the precession will dampen and $M(t)$ will eventually spiral back to its equilibrium state. Equation (2) contains the essence of optical magnetometry operation in which it may be seen that, by determining $f_0$ through some means, an unknown $H_e$ can be directly computed [7,8]. While such an approach is not readily extended suitably for receiving antenna operation, (2) also allows the correct presumption to be made that a time varying effective field magnitude, say due to an additional magnetic field $h_s(t)$ contributing to the effective field such that $H_e(t) = H_0 + h_s(t)$, will result in a time varying precession frequency. This nonlinear behavior, termed a resonant precession modulation (RPM), is visualized in Fig. 1c and is the key to achieving the desired electromagnetic wave magnetic field characterization. Framed in more practical terms, consider the precession of $M(t)$ for a ferrite biased with a sufficiently large magnetic field $H_0$. An electromagnetic wave incident on the ferrite with wavelength sufficiently large such that its magnetic field at the material $h_s(t)$ is approximately spatially uniform, and polarized such that $h_s(t)$ is parallel to $H_0$, will result in a time varying $H_e(t)$. The wave thus manifests as a frequency modulation of the precession, a detection of which then allows for an ensuing extraction of all the information that can be known about the wave.

**Development of analytical expressions.** A theoretical analysis of RPM details explicitly how incident electromagnetic wave information is infused in magnetization dynamics, provides the foundation for which advantages of receiving antenna operation based on RPM may be illuminated, and generates expressions that are verifiable through simulation. Consider the scenario visualized in Fig. 2a in which a ferrite material is biased with a static magnetic field $H_0\hat{z}$ and an electromagnetic wave with magnetic field $h_s(t)\hat{z}$ at the material is incident. As done previously, the assumptions are made that both $H_0$ and the wavelength of the incident wave are sufficiently large. Unlike done previously however, magnetic loss will not be neglected and so a perturbing magnetic field $H_p(t)\hat{x}$ at the frequency $f_0$ defined in (2) exists to maintain magnetization precession with a constant angle θ from its equilibrium state amid the effects of damping. This perturbing field is assumed to be sufficiently small such as to not contribute significantly to $H_e(t)$. Then $M_t(t)$, a component of $M(t)$ transverse to the bias field, may be written from (1) as

$$M_t(t) = M_s sin(\theta) \cos\left(2\pi f_0 t + \int_{t-\tau}^{t} \gamma\mu_0 h_s(t) dt\right) \tag{3}$$

where $\tau = (4\pi f_0 \alpha)^{-1}$ is the characteristic decay time of resonant precession [29,30]. Describing the magnetic field of the incident wave as $h_s(t) = H_s cos(2\pi f_s t + \phi)$, (3) can be evaluated to yield

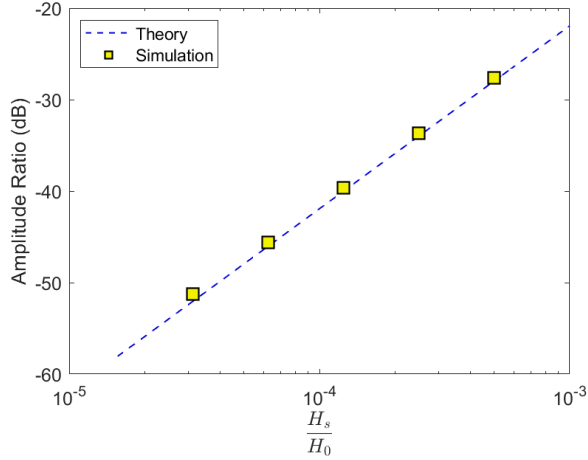

**Fig. 3. Comparison between theory and simulation of RPM operation.** Plot of the transverse magnetization amplitude ratio for a component at the sum or difference frequency $f_0 \pm f_s$ to a component at the unmodulated precession frequency $f_0$. The blue dashed line represents results computed using (8) whereas the yellow squares represent results obtained from micromagnetic simulation.

$$M_t(t) = M_s \sin(\theta) \left[ \cos(2\pi f_0 t) - \frac{1}{2}\gamma\mu_0 H_s \tau \sin(2\pi(f_0 + f_s)t + \phi) \right. \quad (4)$$
$$\left. - \frac{1}{2}\gamma\mu_0 H_s \tau \sin(2\pi(f_0 - f_s)t - \phi) \right]$$

under the conditions of $\gamma\mu_0 H_s \tau \ll 1$ and $2\pi f_s \tau \ll 1$. It may be seen from this equation that the effect of the wave is to generate terms in the precession dynamics at the sum and difference frequencies $f_0 \pm f_s$, and that a detection of $M_t(t)$ theoretically allows all the information about the wave to be extracted from these terms.

Equation (4) can be applied to analyze the amplification achieved by receiving antennas based on RPM as compared to those operating under conventional principles. In order to realize the necessary transduction to an electrical signal, an inductive detection of $M_t(t)$ is assumed using a simple conductive coil of $N$ turns and area $A$ oriented as in Fig. 2a. The amplitude $V_{RPM}$ of the open circuit voltage induced across the terminals of the coil, effectively representative of the strength of the transduced signal, by a term of $M_t(t)$ at $f_0 \pm f_s$ can be found using the principles of Faraday induction [31] to be

$$V_{RPM} = \pi N A \mu_0^2 M_s \sin(\theta) \gamma H_s \tau (f_0 \pm f_s). \quad (5)$$

The same coil operating under conventional antenna principles is visualized in Fig. 2b where the open circuit voltage amplitude $V_{con}$ across its terminals is now induced directly by the incident wave and so, in order to achieve the largest possible voltage amplitude, the coil is oriented with its axis parallel to the magnetic field of the wave. Again applying the principles of Faraday induction, this amplitude is given by

$$V_{con} = 2\pi N A \mu_0 H_s f_s. \quad (6)$$

Taking the ratio of (5) to (6), the signal amplification achieved by a receiving antenna based on RPM as opposed to one operating under conventional principles is found to be

$$\frac{V_{RPM}}{V_{con}} = \frac{1}{2}\left(\frac{M_s \sin(\theta)}{\Delta H}\right)\left(\frac{f_0 \pm f_s}{f_s}\right) \quad (7)$$

where $\Delta H = (\mu_0 \gamma \tau)^{-1}$ is the linewidth of resonant precession [29].

Equation (4) can be further applied to derive a metric independent of precession angle θ that will later be used to evaluate the validity of the developed theory through a comparison with simulation results. Specifically, this metric is the amplitude ratio of the $M_t(t)$ term at either the sum or difference frequency $f_0 \pm f_s$ to that at the unmodulated precession frequency $f_0$, expressed analytically as

$$\left|\frac{M_t^{f_0 \pm f_s}}{M_t^{f_0}}\right| = \frac{1}{2}\frac{H_s}{H_0}Q \quad (8)$$

where $Q = 2\pi f_0 \tau$ is the quality factor [30] of the resonant precession.

Lastly, an expression may be developed to describe the sensitivity δH of receiving antennas based on RPM. In this regard, there exist two practically significant sources of noise that are potentially constraining; the first is noise generated by the inductive detection coil and the second is noise generated by the system instrumentation. With appropriate material choice, inductive detector design, and operating conditions to maximize the value of (5), the noise generated by the coil can be assumed to be non-limiting. Sensitivity is thus dependent on the noise generated by the system instrumentation, in particular, the phase noise $L(f_s)$ associated with producing the perturbing magnetic field at a frequency offset of $f_s$ from the perturbing frequency $f_0$. For detectable electromagnetic wave magnetic fields, $L(f_s)$ must be sufficiently small, satisfying [32]

$$\sqrt{L(f_s)} < \left| \frac{M_t^{f_0 \pm f_s}}{M_t^{f_0}} \right| \qquad (9)$$

where $L(f_s)$ is in linear scale with units of 1/Hz and $M_t^{f_0 \pm f_s}$ is interpreted with units of emu/cm³ Hz$^{1/2}$. Sensitivity can then be expressed using (8) and (9) as [30]

$$\delta H = \frac{2H_0}{Q} \sqrt{L(f_s)} \qquad (10)$$

where $\delta H$ has units of Oe/Hz$^{1/2}$.

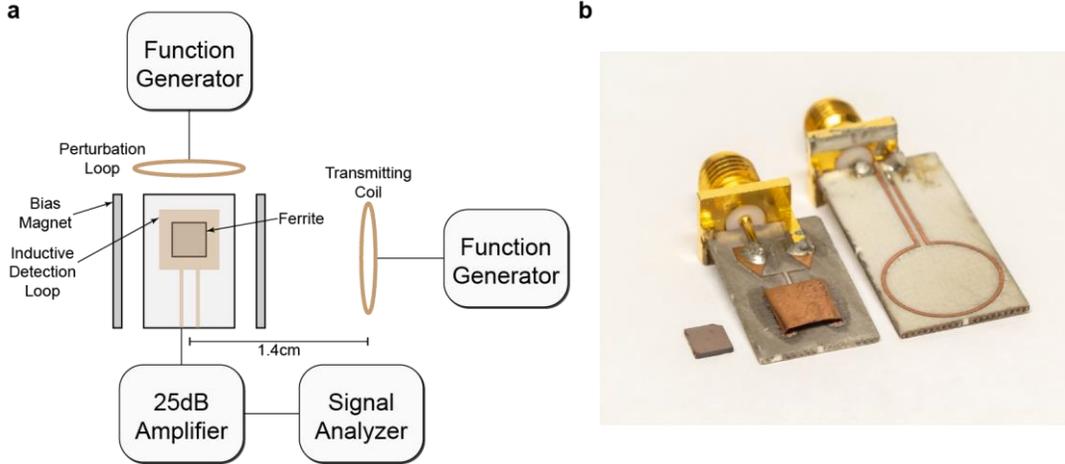

**Fig. 4. RPM based receiving antenna implementation. a** Schematic of the experimental setup indicating placement and orientation of primary components along with connections to test and measurement equipment. A top view looking down onto the plane of the thin film ferrite is taken. **b** Subset of the hardware used for implementation. From left to right: thin film yttrium iron garnet on a gadolinium gallium garnet substrate, loop used for inductive detection of ferrite magnetization dynamics, loop used to produce the perturbation field and maintain magnetization precession.

**Advantages over conventional antenna operation.** The equations derived through theoretical analysis reveal a wealth of information regarding the origin and nature of the advantages of receiving antenna operation based on RPM as well as provide a means of quantifying some of these advantages. For example, equation (7) allows for the identification of two distinct sources from which transduced signal amplification may arise. The first of these sources, yielding the product term $\frac{1}{2}\left(\frac{M_s \sin(\theta)}{\Delta H}\right)$ in (7), is the coupling of magnetization magnetic flux to the inductive detector, where amplification is achieved if this flux is larger than that which may be coupled directly due to the incident electromagnetic wave. The second source, yielding the product term $\left(\frac{f_0 \pm f_s}{f_s}\right)$ in (7), is the generation of magnetization terms of interest at $f_0 \pm f_s$ as a result of frequency mixing. With these frequencies typically being much larger than the electromagnetic wave frequency $f_s$, it follows immediately from the principles of Faraday induction that larger transduced signals will be produced, all else being equal. As a whole, amplification can be more naturally understood from the interpretation of RPM as a method of parametric amplification. In this context, magnetization is the output of a resonant system dependent on the parameter $\boldsymbol{H}_e(t)$, where the perturbing magnetic field $H_p(t)$ behaves as the pump that harmonically drives the system and the electromagnetic wave magnetic field $\boldsymbol{h}_s(t)$ serves to vary the system parameter such that parametric amplification ensues. It is finally noted that the sources of amplification have a remarkable elegance, originating directly from the magnetic material and intrinsic nonlinear nature of its magnetization dynamics, thus requiring no external equipment or circuitry to achieve. For an approximate quantification of attainable values, consider the commonly used yttrium iron garnet (YIG) ferrite material [24,33] biased such that $f_0$ is, say, 1GHz. At this frequency, it is reasonable to expect the linewidth $\Delta H$ of the material to be around 0.2Oe [34]. Then, for the optimistic upper limit case of a highly perturbed, or in other words strongly pumped, magnetization with θ of 90°, signal amplification by a factor upwards of 4000 is reached due solely to the coupling of magnetization magnetic flux. The total transduced signal amplification taking into account frequency mixing as well is then even larger and depends specifically on $f_s$.

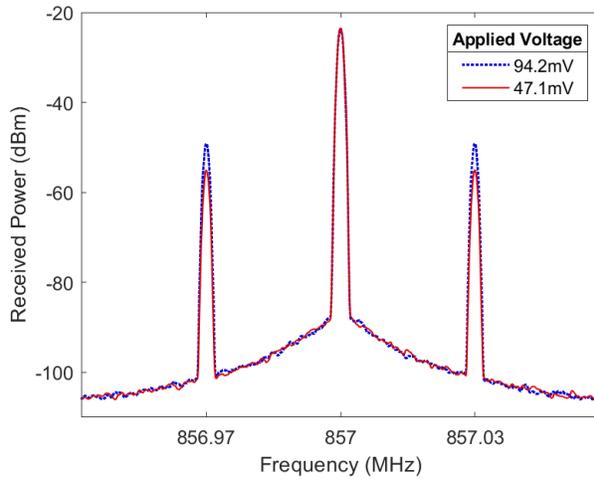

**Fig 5. Experimental received power spectrum.** Spectrum of power coupled to the inductive detection loop. The red solid line corresponds to driving the transmitting coil with a voltage of 47.1mV and the blue dotted line corresponds to driving the coil with a voltage of 94.2mV. In both cases, the coil is driven at a frequency of 30kHz.

**Table I. Amplitude Ratio (dB)**

|  | Voltage (mV) | Experiment | Theory |
|---|---|---|---|
| 10kHz: | 47.1 | -22.32 | -22.47 |
|  | 94.2 | -16.54 | -16.48 |
|  | Voltage (mV) | Experiment | Theory |
| 30kHz: | 47.1 | -31.53 | -31.32 |
|  | 94.2 | -25.60 | -25.34 |
|  | Voltage (mV) | Experiment | Theory |
| 50kHz: | 47.1 | -36.21 | -35.73 |
|  | 94.2 | -30.27 | -29.73 |

**Table II. Amplification (dB)**

| Voltage (mV) | 10kHz | 30kHz | 50kHz |
|---|---|---|---|
| 47.1 | +43.07 | +36.80 | +32.02 |
| 94.2 | +42.70 | +36.49 | +32.86 |

Intimately related to the amplification advantage of receiving antennas based on RPM is their advantage of a physically compact size, where size may be characterized by the size of the inductive detection coil. Simply put, equation (5) indicates that the area $A$ of the inductive detection coil may be reduced by at most the value of (7) while still maintaining a transduced signal strength larger than or equal to that of a conventional receiving antenna of the original size $A$.

Also closely related to the amplification advantage is the sensitivity advantage of receiving antennas based on RPM. Again, the sources of amplification originate directly from the magnetic material and its intrinsic nonlinear magnetization dynamics. This allows for avoidance of the use of external circuitry, and thus transduced signal amplification is achieved without raising the noise generated by the inductive detection coil, immediately indicating a much improved sensitivity as compared to antennas operating under conventional principles. As mentioned previously, the practical limit to sensitivity is then ultimately determined by the instrumentation phase noise. For an approximate quantification of this limit, an ovenized crystal oscillator [35] is considered, producing a 100MHz perturbing field and possessing a phase noise of -170dBc/Hz for $f_s$ of 10kHz. Using a YIG ferrite material biased appropriately with a 35.7Oe field and assumed to have a Q of 2000 [34], (10) then indicates a sensitivity of 0.11nOe/Hz$^{1/2}$, or 11.3fT/Hz$^{1/2}$. This sensitivity is extremely high, comparable to that of modern optical magnetometers [9].

Receiving antennas based on RPM have the additional feature of electronic configurability. From (5) and (7), it may be seen that not only do the perturbing and bias magnetic fields provide degrees of freedom for controlling the signal strength and amplification of the RPM based antenna, but also the bias magnetic field determines the frequency $f_0 \pm f_s$ of the magnetization term to be detected. The benefit of this configurability is that compatibility is readily maintained between the receiving antenna and any external equipment, which may have certain input power or bandwidth limitations, for operation involving electromagnetic waves with a widely varying range of amplitudes and frequencies. In contrast, (6) indicates that an analogous configurability for a conventional receiving antenna is in no way possible.

Of some merit is an explicit distinction between the advantages of RPM based receiving antennas and those of the previously mentioned past antennas incorporating ferrites. Among these past antennas, the ferrite rod antenna [36] is of particular interest due to its relative success and widespread usage [36] as well as its architecture, which closely resembles that of the RPM based receiving antenna. The ferrite rod antenna consists of a conducting coil wrapped around a ferrite core, as illustrated in Fig. 2c, and operates entirely based upon conventional direct voltage induction. The ferrite core serves solely as a source of magnetization to magnify the magnetic flux that an incident electromagnetic wave couples to the coil, likewise magnifying the transduced signal strength. In this regard, the ferrite rod antenna possesses the same advantages of amplification, size, and sensitivity compared to a conventional receiving antenna as does the RPM based receiving antenna. However, not only is the typical amplification achieved by a ferrite rod antenna quite low, falling in the range of approximately a factor of 50 [15], but also the antenna is associated with significant magnetic losses that do not affect the RPM based antenna. These spawn from the fact that its ferrite core is unbiased and include hysteresis loss [37], loss due to domain wall resonance [38], and loss due to thermally activated domain wall movements [36]. The lower amplification of the ferrite rod antenna immediately indicates lower sensitivity as well as lower potential size reduction as compared to the RPM based antenna. The ferrite rod antenna further does not have any means of electronic configurability.

**Considerations for practical implementation.** For the purposes of tractability, a simplified scenario was maintained throughout much of the prior analyses and developments by making various assumptions. While most of these assumptions are easily maintained in practice, a particular few necessitate a thoughtful approach to implementation and consideration of phenomena previously neglected in order to hold. Chief amongst these phenomena is demagnetization. Demagnetization is an equivalent description of the dipole-dipole interaction between subatomic magnetic moments of the ferrite material and is in general dependent on the material shape, size, and magnetization orientation [25]. Accounting for its effects on magnetization dynamics may be accomplished with an additive term $H_d(r,t)$ in the effective field of (1) [28], written now as

$$H_e(r,t) = H_0 + h_s(t) + H_d(r,t) \qquad (11)$$

where $H_d(r,t)$ is known as the demagnetization field and both $H_0$ and $h_s(t)$ are the familiar magnetic fields from prior analyses. Most notable about the demagnetization field in this context is the fact that it is in general spatially nonuniform, thus clearly invalidating the original assumption of a uniform $H_e(r,t)$ and giving rise to complications in the practical implementation of RPM. The ramifications this nonuniformity can be approximately understood as a spatially nonuniform magnetization precession resonance frequency which not only destroys the phase coherence of $M(r,t)$ precession throughout the ferrite, but also introduces additional complicating phenomena such as exchange coupling [25]. The resulting information that may be extracted from a detection of precession during RPM operation is thus significantly degraded in quality if not completely unusable.

The implementation approach taken to address the effects of demagnetization and maintain a uniform effective field involves an appropriate choice of ferrite shape. Specifically, the shape is selected such that, for all orientations of a uniform $M(t)$ occurring throughout normal RPM operation, $H_d(r,t)$ is either negligible in magnitude or approximately uniform in space. With several viable options that may satisfy these conditions, we choose to use ferrites with a thin film geometry for which in-plane orientations of $M(t)$ will yield a negligible demagnetization field whereas out of plane orientations will yield an approximately uniform demagnetization field [20]. Any arbitrary orientation of $M(t)$ is effectively a superposition of an in-plane and an out of plane component and thus it is clear that the uniformity of $H_e(r,t)$ can always be maintained. Ferrites of a thin film geometry, as compared to those of other potentially suitable shapes, have the additional benefit that associated processing methods such as pulsed laser deposition, spin spray plating, or liquid phase epitaxy are mature enough to produce films of extremely high quality [21].

With the complications accompanying demagnetization accounted for, all other details of practical implementation follow in a fairly straightforward manner. In this work, the bias field $H_0$, required to establish uniformity of $M(t)$ as well as a precession center frequency about which the modulation of RPM occurs, is applied with permanent magnets. The perturbation field $H_p(t)$, required to maintain $M(t)$ precession amid a finite magnetic loss, is applied using a conducting coil driven continuously at $f_0$ of (2). Lastly, inductive detection of magnetization dynamics, required to obtain an electrical signal from which information may be extracted, is performed with another conducting coil. With regards to this detection, it of interest to recognize that, since $H_d(r,t)$ is not always negligible throughout operation of RPM using a thin film ferrite, (11) indicates that it still has some effect on the magnetization precession dynamics. This effect is a modification of the precession such that it approximately resembles a pendulum oscillation [39] as visualized in Fig. 1d. While this has no significant bearing on operation, it is of importance in deciding upon orientation of the inductive detector to achieve the largest transduced signal.

**Micromagnetic simulation.** As a means of evaluating the validity of the RPM concept, the analytical expressions describing its operation, and the approach to its practical implementation, micromagnetic simulations are performed. Specifically, simulations are carried out using the Object Oriented Micromagnetic Framework (OOMMF) [40], a micromagnetic simulator widely used and well recognized as the standard for accurate solutions [27]. Reproducing the scenario visualized in Fig. 2a, a ferrite material is biased with a magnetic field $H_0$ of magnitude 50Oe, resulting in a precession resonance frequency $f_0$ of 848MHz. An electromagnetic wave magnetic field $h_s(t)$ is applied as a sinusoid with various amplitudes at a frequency of 50kHz. Lastly, a perturbing magnetic field $H_p(t)$ is applied as a sinusoid with amplitude 0.025Oe at a frequency of 848MHz. It should be noted that the amplitude of $H_p(t)$ is somewhat arbitrary and was need only be sufficiently small such that it does not contribute significantly to the effective field. The amplitudes and frequency of $\boldsymbol{h_s}(t)$ are similarly arbitrary to an extent and need only satisfy the conditions for which (4) was derived. In accordance with the approach to implementation taken to account for the effects of demagnetization, the ferrite is modeled to have a thin film geometry of dimensions 1x1x0.001mm, with the material itself having the properties of YIG and a Gilbert damping coefficient α of 1e-3.

Simulation is performed for a total time of 0.2 milliseconds for each $\boldsymbol{h_s}(t)$ amplitude of interest from which the transverse magnetization component $M_t(t)$ is extracted and the ratio of $M_t^{f_0 \pm f_s}$ to $M_t^{f_0}$ is determined. This ratio is also computed theoretically from (8) using

$$Q = \frac{\sqrt{H_0(H_0 + M_s)}}{\alpha(2H_0 + M_s)}, \tag{12}$$

the quality factor of unmodulated precession for a thin film ferrite [39], and a comparison with the simulation results is plotted in Fig. 3. The simulations not only demonstrate that the concept of RPM is feasible and support the proposed approach to deal with the effects of demagnetization, but also the excellent agreement that may be seen from Fig. 3 validates the developed analytical expressions.

**Experimental implementation.** A receiving antenna operating based on RPM is implemented using available hardware and materials. Results from experiments using the antenna provide further evidence supporting the feasibility of RPM based operation, validate the practical applicability of the developed RPM theory, and are already able to demonstrate significant advantages over conventional receiving antenna operation. The experimental setup is overviewed in Fig. 4a and employs a YIG thin film ferrite of dimensions 3.5x5x0.003mm epitaxially grown on a gadolinium gallium garnet (GGG) substrate. Inductive detection of the transverse component of magnetization $M_t(t)$ is accomplished with a single turn loop of area 6mm$^2$ constructed from a piece of copper soldered onto traces fabricated on a Rogers 4003C board. This loop is connected to a signal analyzer via a 25dB amplifier, with the amplifier used to ensure observable signal levels above the noise floor throughout extensive testing. Two neodymium permanent magnets are used to apply the bias magnetic field $\boldsymbol{H_0}$ with a magnitude of approximately 50Oe, corresponding to a measured precession resonance frequency of 857MHz. The field $\boldsymbol{h_s}(t)$ is produced by an electromagnetic wave transmitted from a 125-turn coil of area 3375mm$^2$ positioned 1.4cm from the center of the inductive detection loop and driven by a function generator. This 1.4cm distance is chosen to correspond to approximately ten times the effective radius of the inductive detection loop. Lastly, the perturbation field $H_p(t)$ is applied with a single turn loop of area 95mm$^2$ fabricated on a Rogers 4003C board and driven by a function generator. For all results from this setup, the perturbation loop was driven sinusoidally with a fixed input power of -25dBm at the precession resonance frequency 857MHz whereas the transmitting coil was driven sinusoidally with varied voltages and frequencies to evaluate trends in the behavior of the system. The YIG film, inductive detector, and perturbation loop are displayed in Fig. 4b.

For the purposes of comparison and assessment of the performance of the RPM based receiving antenna, the inductive detection loop of the RPM implementation was also operated under conventional receiving antenna principles in a second experimental setup. This setup is essentially visualized in Fig. 2b and involves the same electromagnetic waves as in the first experimental setup now directly inducing voltages in the inductive detection loop in order to achieve transduction into electrical signals. The electromagnetic waves are transmitted from the transmitting coil of the first experimental setup driven sinusoidally by a function generator with the same voltages and frequencies as previously. The inductive detection loop was connected directly to a signal analyzer, as the 25dB amplifier was not compatible for the frequencies used, and the same 1.4cm distance separated it from the transmitting coil.

From the RPM implementation of the first setup, the power coupled to the inductive detection loop is visualized in Fig. 5 for the cases of the transmitting coil driven with voltages of 47.1mV and 94.2mV across its terminals at a frequency of 30kHz. In both cases, the expected sum and difference frequency terms $f_0 \pm f_s$ as well as the unmodulated frequency term $f_0$ are clearly evident, demonstrating that RPM operation is achieved. Comparing between the two cases then it is seen that doubling the driving voltage effectively doubles the amplitude ratio of (8). Recognizing that the driving voltage is directly proportional to the $\boldsymbol{h_s}(t)$ field amplitude $H_s$ by antenna theory [15,16], then the linear relationship between the amplitude ratio and $H_s$ is thus confirmed. Table I presents additional detailed data of this ratio, obtained driving the transmitting coil with various combinations of voltages and frequencies, and provides a comparison with the theoretical expectations. The theoretical expectations were computed using (8) where $H_0$, $H_s$, and $Q$ were determined through procedures detailed in the Methods section. This table shows outstanding agreement between the computed values and those experimentally observed, solidifying the validity of the developed RPM theory as well as its applicability to a practical implementation.

To assess the performance of the RPM based antenna, an amplification metric is computed by comparing the power coupled to the inductive detection loop in the first setup to that in the second setup, where the loop operates as a receiving antenna based on conventional principles. This data is presented in Table II and, again, is obtained driving the transmitting coil with various combinations of voltages and frequencies. From this table, it is seen that significant amplification is achieved for all cases under consideration, demonstrating promise for RPM based antenna application in the very low frequency (VLF) and low frequency (LF) bands. It is also possible to conclude from Table II that amplification grows as the transmitting coil driving frequency decreases and that amplification is also more or less invariant with respect to the transmitting coil driving voltage, both expected behaviors based on (7).

## DISCUSSION

Receiving antennas based on RPM are introduced and their principal advantages in amplification, sensitivity, and size as compared to receiving antennas operating under conventional principles are demonstrated. Not only is the general concept of RPM operation proven to be viable through simulation and experiment, but also its benefits are clearly exhibited in experiment for reception of electromagnetic waves with frequencies in the VLF and LF bands. With the RPM based receiving antenna implementation in this work being of a rather makeshift nature, it is expected that further refinements will only establish the practical capabilities of these antennas more substantially. For example, an inductive detector design that mitigates shielding effects on the ferrite material will result in a larger transduced signal as would an increase in the ferrite material volume. Most promising about RPM based antenna operation is the fact that it is fundamentally different from conventional antenna operation, possessing the potential to contribute a completely new paradigm to antenna theory and overcome longstanding barriers in performance.

## METHODS

**Theoretical amplitude ratio.** The theoretical amplitude ratios of Table I were computed using (8). The value of $H_0$ was found with a DC gaussmeter. The values of $H_s$ were found with a setup involving the perturbation loop of the RPM antenna implementation used as a conventional receiving antenna. Specifically, the loop was connected directly to a signal analyzer and positioned 1.4cm from the transmitting coil of the RPM antenna implementation. The transmitting coil was driven with the voltages and frequencies of consideration in Table I by a function generator and the principles of Faraday induction were applied to the reception results to extract $H_s$ values. The value of $Q$ was found with a circuit model for the RPM antenna displayed in Supplementary Fig. 1. The value for each circuit element was obtained through curve fitting of the circuit model impedance spectrum to the measured RPM antenna impedance spectrum. Spectrum measurement was accomplished by connecting the RPM antenna to a network analyzer and imposing on it the same biasing and perturbation power conditions under which the experimental values of Table I were found.

# Supplementary Information

Kevin Quy Tanh Luong, Wei Gu, Foad Fereidoony, Lap Yeung, Zhi Yao & Yuanxun Ethan Wang

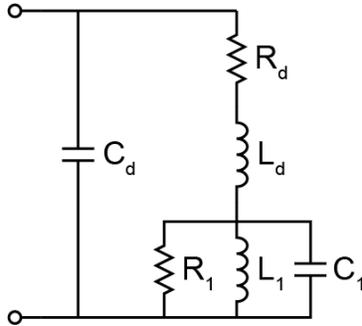

**Supplementary Fig. 1. RPM based receiving antenna circuit model.** Schematic of circuit model representing the RPM based receiving antenna where $R_d$, $C_d$, and $L_d$ are dependent on the inductive detector whereas $R_1$, $L_1$, and $C_1$ are dependent on the ferrite material.